\documentclass[conference]{IEEEtran}
\IEEEoverridecommandlockouts
\usepackage{spconf}
\usepackage{algorithm,graphicx}
\usepackage{algorithmic}
\usepackage{url}
\usepackage{cite}
\usepackage[usenames]{color}
\usepackage{amsfonts}
\usepackage{fancyhdr}
\usepackage{arydshln}
\usepackage{amsmath,amssymb,amsthm}%
\usepackage{graphicx,psfrag,caption,subcaption}
\usepackage{listings}%
\usepackage{arydshln}
\usepackage{algorithmic}%
\input{mysymbol.sty}
\DeclareMathOperator*{\argmaxA}{arg\,max} % Jan Hlavacek

\title{Fast Graph Convolutional Recurrent Neural Networks}
\name{Sai Kiran Kadambari and  Sundeep Prabhakar Chepuri 
\thanks{This work is supported by the Tata Trusts.}}
\address{Indian Institute of Science, Bangalore, India\\
Email:~\{kadambarik;spchepuri\}@iisc.ac.in}

\begin{document}
%\ninept
%
\maketitle
\begin{abstract}
	\textbf{
This paper proposes a Fast Graph Convolutional Neural Network (FGRNN) architecture to predict sequences with an underlying graph structure.
The proposed architecture addresses the limitations of the standard recurrent neural network (RNN), namely, vanishing and exploding gradients, causing numerical instabilities during training.
State-of-the-art architectures that combine gated RNN architectures, such as Long Short-Term Memory (LSTM) and Gated Recurrent Unit (GRU) with graph convolutions are known to improve the numerical stability during the training phase, but at the expense of the model size involving a large number of training parameters.
FGRNN addresses this problem by adding a weighted residual connection with only two extra training parameters as compared to the standard RNN.
Numerical experiments on the real 3D point cloud dataset corroborates the proposed architecture.
	}
\end{abstract}

\vspace*{-1mm}
\maketitle

\section{Introduction}
Many real-world datasets occur as structured sequences, e.g., space-time series, or graph/network-time series. 
Typical examples of such data are multichannel speech data, videos, dynamic point clouds, timeseries data from brain networks or social networks, where the successive frames might have a dynamic pattern and each time frame might have a spatial or graph structure. 
Developing generative models to process such structured time-varying data for prediction or interpolation is of significant interest in data analytics.

Convolutional Neural Network (CNN) and Recurrent Neural Network (RNN) are two popular variants of neural networks commonly used in a wide variety of engineering and science applications. CNN comprises of a convolutional layer and pooling layer as hidden layers and is known to capture intricate structures in the data. 
Whereas, RNNs have a memory element that captures the information about the past and decisions are based on the gathered memory.
%RNNs are typically used for sequence learning, e.g., in natural language processing.  

Architectures combining RNN and CNN have been proposed to find patterns in time-varying data~\cite{Donahue:2017:LRC:3069214.3069251, Karpathy, Vinyals_2015}.
These models assume that the data is defined on a regular domain, and the convolutions in this case are simple 2-D convolutions.
%for example, an image can be considered as data indexed by a regular two-dimensional grid, and the convolutions, in this case, are simple 2-D convolutions. 
In many cases, the data available might not be supported on a regular domain~\cite{GeometricDeepLearning,monti2017geometric}. Examples of data residing on irregular domains are data from weather monitoring stations, biological networks, and social networks, to list a few.
Graphs may be used to represent the irregular domain on which the data is defined and explain the complex relationships in such data~\cite{Bigdata,sandryhaila2014discrete}. 
More specifically, the data is indexed by the nodes of the graph, and the edges encode the pairwise relationships between the nodes. Numerous approaches have been proposed to learn the underlying graph structure from the available data ~\cite{LearnGraphData,chepuri2016learning}.

The standard RNN architecture is known to have vanishing or exploding gradients, which makes it unstable during training. A particular class of RNN that overcomes this problem is the Long Short-Term Memory (LSTM) architecture~\cite{hochreiter1997long}. These variants of RNNs have a gated architecture with a large number of training parameters. As the number of trainable parameters in LSTM is large, they take more training time, even on high-end computational machines, due to the required memory.

%Generalizing CNNs for the data defined on irregular domains, for the problems like node classification and link prediction, 
Graph Convolutional Neural networks (GCNN) that generalize the CNNs to handle graph-structured data on arbitrary graphs have been proposed in~\cite{defferrard2016convolutional,ThomasKipf}, where the convolution operator is now generalized using polynomials of the graph Laplacian matrix.
When the graph data is time-varying (e.g., dynamic 3D point cloud data), to capture the temporal variations of such data, LSTM combined with GCNN is proposed in~\cite{Seo_2018}, where GCNNs capture the spatial structure and LSTMs capture the temporal variations of the data.

In this paper, we propose a stable architecture of RNN that combines GCNN with a standard RNN. To stabilize the gradients a weighted residual connection is introduced.
% of the standard RNN for the task of sequence modeling. 
The proposed architecture has a much fewer number of training parameters as compared to architectures that combine LSTM with GCNN. 
Hence the term fast in FGRNN. The proposed architecture is inspired by~\cite{kusupati2018fastgrnn}, where they propose a stable, scalable, and a faster variant of RNN for data defined on regular domains.
The main contributions of this paper are as follows. We propose a stable FGRNN architecture for efficient training and prediction of the data defined on irregular domains.
We show that the proposed FGRNN architecture with only $2$ additional parameters as compared to the standard RNN is stable during the training phase and overcomes the vanishing or exploding gradients problem. The experiments on real 3D point cloud data, where the task is to predict the next 3D point cloud frame, demonstrate the developed theory.

Throughout this paper, we will use upper and lower case boldface letters to denote matrices and column vectors, respectively.
%We will denote sets using calligraphic letters. ${\bf 1}$(${\bf 0}$) denotes the vector/matrix of all ones (zeros) of appropriate dimension. 
$\diag[\cdot]$ is a diagonal matrix with its argument along the main diagonal.
%${\rm diag}(\bbX)$ is a diagonal matrix with the diagonal elements of $\bbX$ along its main diagonal. 
$X_{ij}$ and $x_i$ denote the $(i,j)$th element and $i$th element of $\bbX$ and $\bbx$, respectively.
\section{Problem Statement}
Consider a graph $\ccalG = (\ccalV,\ccalE)$ with $N$ vertices (nodes), where $\ccalV$  is the set of vertices (or nodes) and $\ccalE$ denotes the edge set such that $(i,j) \in \ccalE$ if there is a connection between node $i$ and node $j$. 
The structure of the graph with $N$ nodes is captured by the weighted adjacency matrix $\bbA \in \reals^{N \times N} $ whose $(i,j)$th entry denotes the weight of the edge between node $i$ and node $j$. 
%When there is a no edge between node $i$ and node $j$, the $(i,j)$th entry of $\bbA$ is zero. 
This matrix represents the network connectivity. We assume that the graph is undirected with positive edge weights. The corresponding graph Laplacian matrix is a symmetric matrix of size $ N $, given by $\bbL = \bbI_N - \bbD^{-1/2}\bbA\bbD^{-1/2}$,
where $\bbD \in \reals^{N \times N}$ is the diagonal matrix whose diagonal entries are given by the length $N$ degree vector $ \bbd = \bbA{\bf 1} $.
We call the set of graph signals $\{\bbx_t \}^T_{t=1}$, indexed by the vertices of the graph $\ccalG$, collected in the $N \times T$ matrix $\bbX = [\bbx_1, \bbx_2,...,\bbx_T]$ as graph data. 
That is, $\bbx_t \in \reals^N$ is the graph signal at time $t$.

The goal of this paper is to learn a function for sequence modeling, where the task is to predict the most likely feature vector based on the previous observations.
More percisely, given the previous $T$ observations of the data, we are interested in predicting the most likely feature vector $\bbx_{t+1}$:
%\vspace{-0.07in}
\begin{equation}\label{eq:prediction_eqn}
\begin{aligned}
& \hat{\bbx}_{t+1} = &&\argmaxA_{\bbx_{t+1}}  \quad P(\bbx_{t+1}|\bbx_{t-T},...,\bbx_t ),
\end{aligned}
\end{equation}
where in \eqref{eq:prediction_eqn} we maximize the likelihood of $\bbx_{t+1}$ given the previous $T$ observations.
Such problems usually appear in language modeling, image prediction, to name a few,  wherein the task is to predict the most likely feature vector given the previous $T$ observations.
In this paper, we focus on the case where the data at each time instance has a structure, which is defined by the graph $\ccalG$.
\section{Graph convolutions}
As the graphs do not have a natural ordering, the standard convolution operation cannot be generalized to arbitrary graphs using localized filters. Hence, a spectral definition of graph convolution is defined in ~\cite{Bruna} and is based on an elementwise multiplication in the graph frequency domain. 
Using this definition, the graph convolution of the graph signal $\bbx$ is given as 
$ \bby = g_\theta \star_\ccalG \bbx = \bbU g_\theta(\boldsymbol{\Lambda})\bbU\bbx$, where $\bbU \in \reals^{N \times N}$ is the matrix of eigenvectors, $\boldsymbol{\Lambda} \in \reals^{N \times N}$ is the matrix of eigenvalues of the graph Laplacian matrix $\bbL$. Here, $g_\theta(\boldsymbol{\Lambda})$ represents the responce of the graph filter in the frequency domain and $ \star_\ccalG $  is the graph convolution operator.  
This method is computationally expensive, as it involves multiplications with a dense eigenvector matrix $\bbU$.
%Moreover, it requires computing eigendecomposition in the first step.
To circumvent this issue,~\cite{defferrard2016convolutional} defined the graph convolution operation by approximating $g_\theta (\boldsymbol{\Lambda})$ with a truncated Chebyshev polynomial expansion of order $K$. Mathematically, the graph convolution operation is given by
\vspace*{-0.10in}
\begin{equation}\label{eq:Conv_1}
\begin{aligned}
g_\theta \star_\ccalG \bbx = \sum_{k=0}^{K-1} \theta_k T_k(\tilde{\bbL})\bbx,
\end{aligned}
\end{equation}
where the parameter $\boldsymbol{\theta} = [\theta_0,\theta_1,...\theta_{K-1}]^{\rmT} \in \reals^K$ is a vector of Chebyshev coefficients and $T_k(\tilde{\bbL}) \in \reals^{N \times N}$ is the Chebyshev polynomial of order $K$ evaluated at the normalized Laplacian matrix defined as $\tilde{\bbL} =2\bbL/\lambda_{max} - \bbI_N$.
The above graph convolution operation incurs linear complexity, i.e., $\ccalO(K|\ccalE|)$, due to the recurrence relation $T_k(x) = 2xT_{k-1}(x)- T_{k-2}(x)$. Here, $T_0 = 1$ and $T_1 = x$. 
%The above convolution operation is equivalent to a graph filter implemented using the $K$ order polynomial of the Laplacian matrix.

A first order approximation $(K=1)$ of the \eqref{eq:Conv_1} is proposed in~\cite{ThomasKipf}.
% for the task of semi-supervised classification with graph data.
Using the approximation in~\cite{ThomasKipf}, the convolution operation simplifies to
\vspace*{-0.10in}
\begin{equation}\label{eq:Conv_2}
\begin{aligned}
&g_\theta \star_\ccalG \bbx  &&= \bbW \tilde{\bbL}_1 \bbx,
\end{aligned}
\end{equation} 
where $\tilde{\bbL}_1 =\bbI_N + \bbD^{-1/2} \bbA \bbD^{-1/2} $, $\bbW \in \reals^{P \times N}$ is the filter parameter and $P$ is the hidden state dimension.
While we focus on the graph convolution framework introduced in~\cite{ThomasKipf} for the analysis of the proposed architecture, we demonstrate the effectiveness of the proposed model using both \eqref{eq:Conv_1} and \eqref{eq:Conv_2}. 

%The number of parameters parameters shared by the whole graph are $2$ and the graph convolution considers the one hop neighbours from the center nodel. 
%To effectively convolve the $K^{th}$ order neighbourhood of a node, we stack $K$ layers of this form.
%We can constrain the number of parameters, for instance $\theta_0 = -\theta_1 = \theta$ the graph convolution operation is given by g$_\theta \star_\ccalG \bbx  = \theta(\bbI_N + \bbD^{-1/2} \bbA \bbD^{-1/2})\bbx $.
%The eigen values of  $\bbI_N + \bbD^{-1/2} \bbA \bbD^{-1/2}$ are in the range $[0,2]$.
%While developing deep models, we stack these layers and repetedly apply this operation will lead to vanishing and exploding gradients which can be made less sever  by renormalization trick. If we assume the data is $\bbX \in \reals^{N \times C}$ with $C$ channels, the convolution operation can be generalized as  
%\begin{equation}\label{eq:Conv_2}
%\begin{aligned}
%&g_\theta \star_\ccalG \bbx  &&= \bbW \tilde{\bbL}_1 \bbx
%\end{aligned}
%\end{equation}
%where $\tilde{\bbL}_1 =\bbI_N + \bbD^{-1/2} \bbA \bbD^{-1/2} $  

\section{Recurrent neural networks}\label{sec:RNNs}
For tasks like sequence modeling, standard RNNs are typically used.
The standard RNN maintains a hidden state that captures the temporal variations in the data. 
The hidden state in the standard RNN at the time instance $t$ is given by $\bbh_t =  \sigma(\bbW\bbx_t + \bbU \bbh_{t-1} + \bbb)$,
%\begin{equation}
%\begin{aligned}
%\bbh_t =  \sigma(\bbW\bbx_t + \bbU \bbh_{t-1} + \bbb),
%\end{aligned}
%\end{equation} 
where $\bbW \in \reals^{P \times N}$, $\bbU \in \reals^{P \times N}$, and $\bbb\in \reals^{P \times 1}$ are the training parameters. Here, $\sigma(\cdot)$ is the nonlinear activation function, and typical choices for $\sigma(\cdot)$ are such as $\rm tanh$, $\rm ReLU $, $\rm sigmoid$. 

It is well known that standard RNNs suffers from exploding and vanishing gradient issues, due to which they are highly unstable during the training phase.  
%Thus more stable variants of RNNs like LSTMs and Gated Recurrent Units (GRU) are used while dealing with time-varying signals. 
%These models have a gated architecture that helps to stabilize the gradients, but they are more expensive to train as the number of training parameters are often large.
A residual connection is introduced in~\cite{kusupati2018fastgrnn} to alleviate this numerical instability. 
This fast variant of RNN in short, (FRNN) has much fewer parameters as compared to LSTM.
Mathematically, the hidden states of FRNN is given by 
\begin{equation}
\begin{aligned}
&\tilde{\bbh}_t&& = \sigma(\bbW\bbx_t + \bbU \bbh_{t-1} + \bbb) \\
&\bbh_t &&=   \alpha \tilde{\bbh}_t + \beta \bbh_{t-1},
\end{aligned}
\end{equation}
where $\alpha$ and $\beta $ are scalar the training parameters. 
%Given the hidden state at time $T$, the next feature is predicted using a linear relation $\bbx_{T+1} = \bbV \bbh_T + \bbz$, where $\bbV \in \reals^{N \times P}$, $\bbz \in \reals^{N \times 1}$ are the training parameters.
FRNN updates the hidden states in a controlled manner using only 2 extra training parameters $\alpha$ and $\beta$.
Moreover, the number of additional computations required per time step is $N$, which is usually a tiny fraction as compared to the operations in RNN and as such it is very small compared to the computational complexities of the gated architectures like LSTM and GRU. 
\vspace*{-0.10in}
\section{Fast graph recurrent neural networks}
In this section, we propose a Fast Graph Convolutional Recurrent Neural Network (FGRNN) architecture that combines FRNN with GCNNfor the task of prediction.
For the task of sequence modeling for data defined on regular domains, architectures that combine RNN and CNN are proposed in~\cite{SHI}.
%Architectures that combine RNNs and CNNs are proposed in~\cite{SHI}, where the task is sequence modeling of data defined on a regular grid domain.
%This can be seen as a particular case of the proposed model, where the graph is an image grid with the nodes being well ordered.
In this architecture, LSTM is used, where the 2D convolutions replace the multiplications with the dense matrices.
Inspired by this approach, when the data is defined on irregular domains, we propose to replace the multiplications by the dense matrices in FRNN with graph convolutions. 
Specifically, the update equations for the hidden state is given by 
\begin{equation}\label{eq:FGRNN_equn}
\begin{aligned}
&\tilde{\bbh}_t&& = \sigma(\bbW \star_\ccalG \bbx_t + \bbU \star_\ccalG \bbh_{t-1} + \bbb) \\
&\bbh_t &&=   \alpha \tilde{\bbh}_t + \beta \bbh_{t-1},
\end{aligned}
\end{equation}
where recall that $\star_\ccalG$ is the graph convolution operator defined in \eqref{eq:Conv_1} or \eqref{eq:Conv_2}. We focus on \eqref{eq:Conv_2}, for the sake of simplicity, for which the hidden state at time $t$ smiplifies to
\begin{equation}\label{FRNN}
\begin{aligned}
&\tilde{\bbh}_t&& = \sigma(\bbW \bbL \bbx_t + \bbU \bbL \bbh_{t-1} + \bbb), \\
&\bbh_t &&=   \alpha\tilde{\bbh}_t + \beta \bbh_{t-1}.
\end{aligned}
\end{equation}

Given the hidden state at any time step $t$, we predict the next feature using the linear relation given by $\hat{\bbx}_{t+1} = \bbV\bbh_t+ \bbz$, where $\bbV \in \reals^{N\times P}$ and  $\bbz \in \reals^{N}$ are the training parameters. 
At each time step, we define the prediction loss function as $J_t(\hat{\bbx}_{t+1},\bbx_{t+1}, \boldsymbol{\Theta}) = \|\bbx_{t+1} - \hat{\bbx}_{t+1} \|^2_2$, where $\boldsymbol{\Theta} = \{ \bbW, \bbU, \bbV, \alpha, \beta, \bbb,\bbz \}$ is the set of all training parameters.
The prediction loss function after $T$ time steps is given by $J_T = \sum_{t=1}^{T} J_t(\bbx_{t+1}, \hat{\bbx}_{t+1},\boldsymbol{\Theta})$.

To update the training parameters during the training phase using backpropagation through time (BPTT), we calculate the gradient of the loss function with respect to training  parameters. For simplicity, assume that $\boldsymbol{\Theta} = \{ \bbW, \bbU, \bbV\}$ are the trainig parameters.
Thus, the gradients of the loss functions $J_T$ w.r.t. parameters $\bbW$, $\bbU$, and $\bbV$ are given by
\begin{equation}\label{gradients_FRNN}
\begin{aligned}
&\frac{\partial J_T}{\partial \bbW} && = \sum_{t=1}^{T} \frac{\partial J_t}{\partial \bbh_T} \prod_{t=2}^{T} \frac{\partial \bbh_t}{\partial \bbh_{t-1}} \frac{\partial \bbh_1}{\partial \bbW},\\
& \frac{\partial J_T}{\partial \bbU}  &&= \sum_{t=1}^{T} \frac{\partial J_t}{\partial \bbh_T} \prod_{t=2}^{T} \frac{\partial \bbh_t}{\partial \bbh_{t-1}} \frac{\partial \bbh_1}{\partial \bbU}, \\
&\frac{\partial J_T}{\partial \bbV} &&= \sum_{t=1}^{T} \frac{\partial J_t}{\partial \bbh_T} \prod_{t=2}^{T} \frac{\partial \bbh_t}{\partial \bbh_{t-1}} \frac{\partial \bbh_1}{\partial \bbV},
\end{aligned}
\end{equation}

where the term $\frac{\partial \bbh_t}{\partial \bbh_{t-1}} = \alpha \bbD_t \bbU \bbL+ \beta \bbI_N \in \reals^{N \times N}$ common to all the gradients determines the numerical stability during the training phase.
%the critical term in the above expression for the gradient.
This term is multiplied by itself $T-2$ times,and depending on the conditioning of $\bbU$ and/or $\bbL$, $ \prod_{t=2}^{T} \frac{\partial \bbh_t}{\partial \bbh_{t-1}}$ may quickly become ill-conditioned with $\beta =0$.
Here, $\bbD_t$ is the diagonal matrix with pointwise nonlinearity given by $\bbD_t = \rm{ diag}(\sigma'(\bbW\bbL\bbx_t + \bbU \bbL\bbh_t + \bbb))$ with $\sigma'(\cdot)$ being the gradient of $\sigma(\cdot)$. 
As the special case, if the activation function is $\rm ReLU$, then the matrix $\bbD_t $ is an identity matrix. 
Notice that when $\alpha = 1$ and $\beta = 0$, we have the standard RNN architecture.

\begin{figure*}[!h]
	\centering
	\psfrag{celcius}{\footnotesize $^\circ$ C}
	\psfrag{5}{\tiny 5} \psfrag{6}{\tiny 6} \psfrag{7}{\tiny 7} \psfrag{8}{\tiny 8} \psfrag{9}{\tiny 9} \psfrag{10}{\tiny 10} \psfrag{11}{\tiny 11}
	\begin{subfigure}[t]{0.20\textwidth}
		\includegraphics[width=1.5in, height=1.5in]{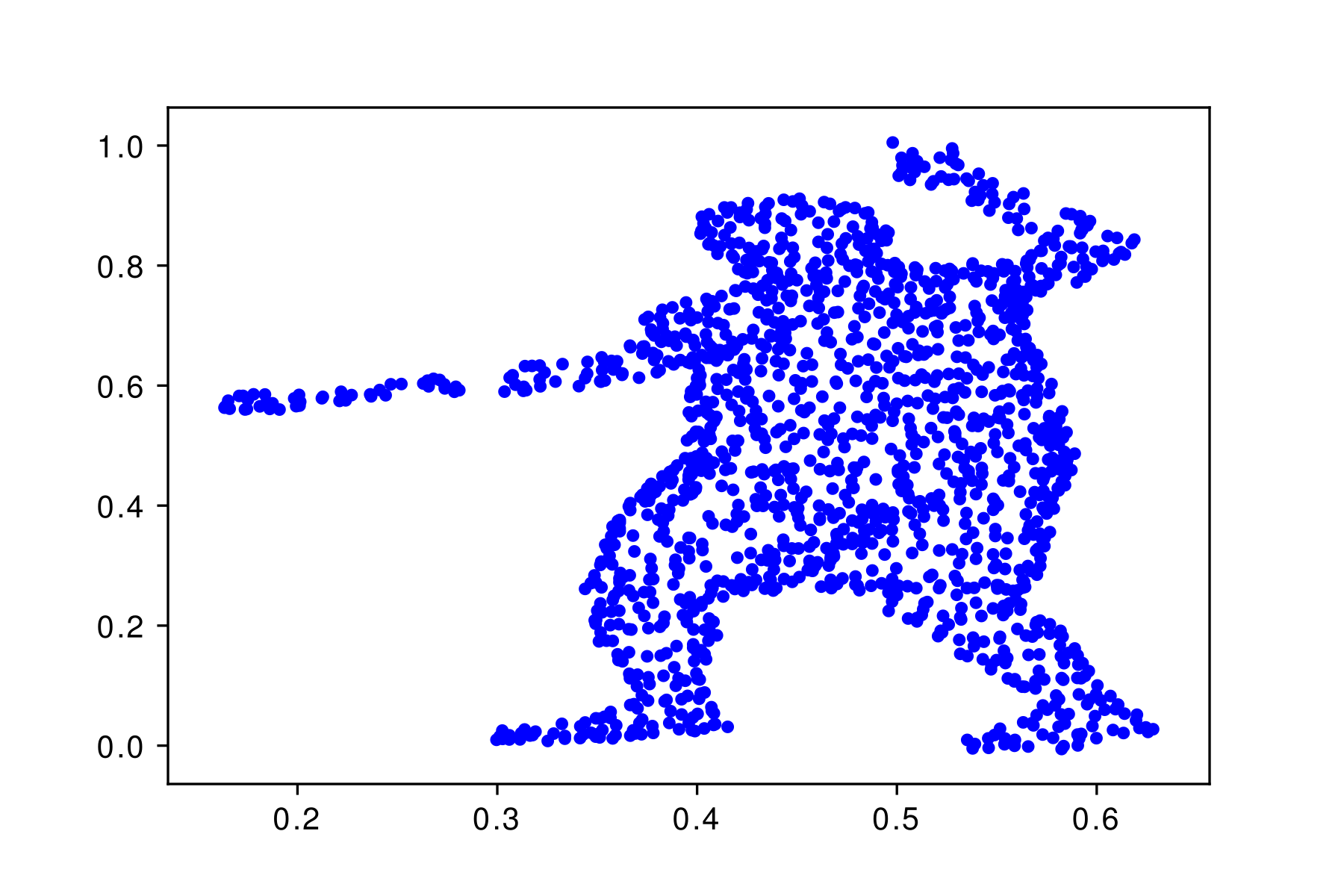}
		\caption{}
		\label{fig:Test_loss_methods}
	\end{subfigure}%
	~
	\begin{subfigure}[t]{0.20\textwidth}
		\includegraphics[width=1.5in, height=1.5in]{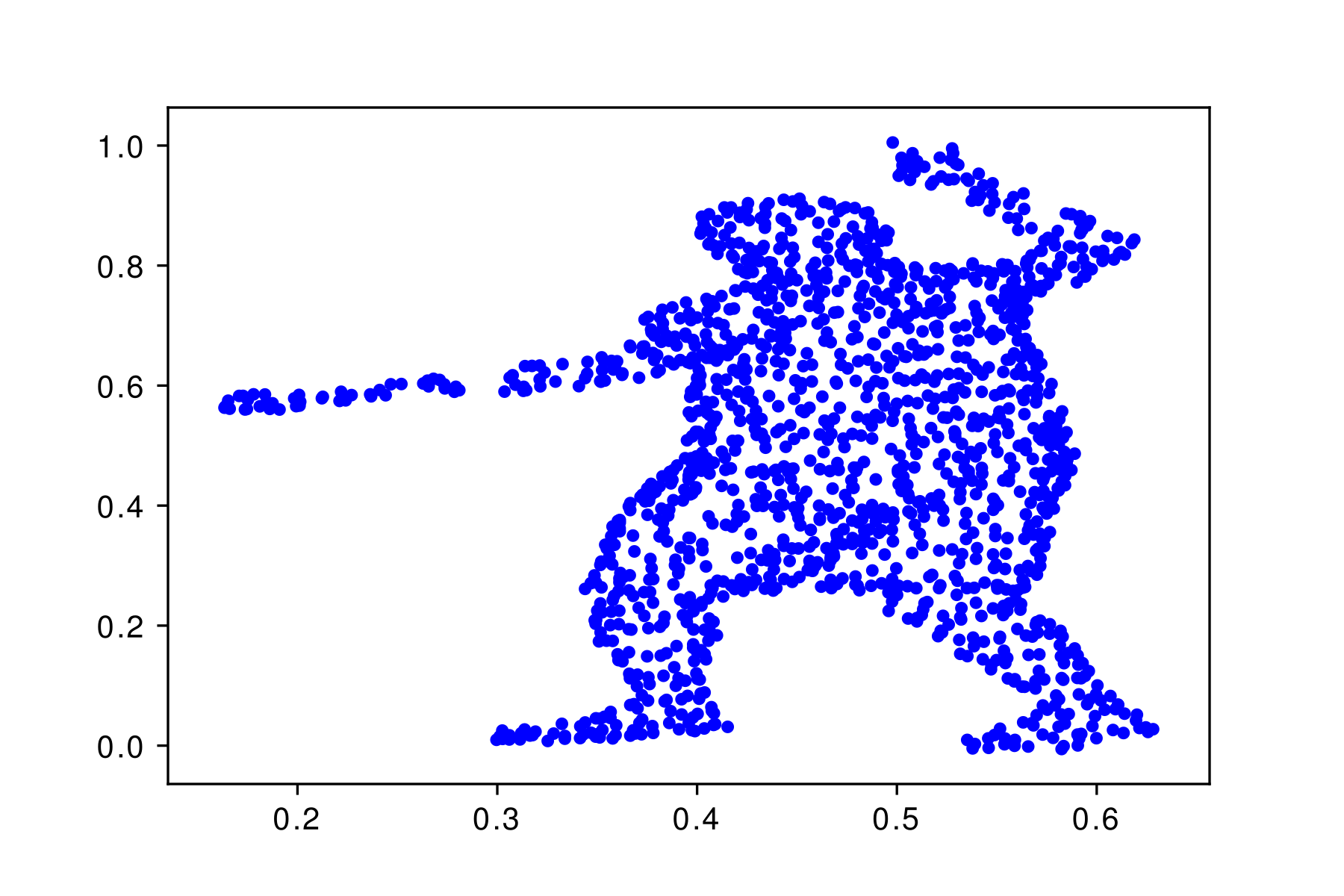}
		\caption{}
		\label{fig:prop_cheb}
	\end{subfigure}%       
	~
	\begin{subfigure}[t]{0.20\textwidth}
		\includegraphics[width=1.5in, height=1.5in]{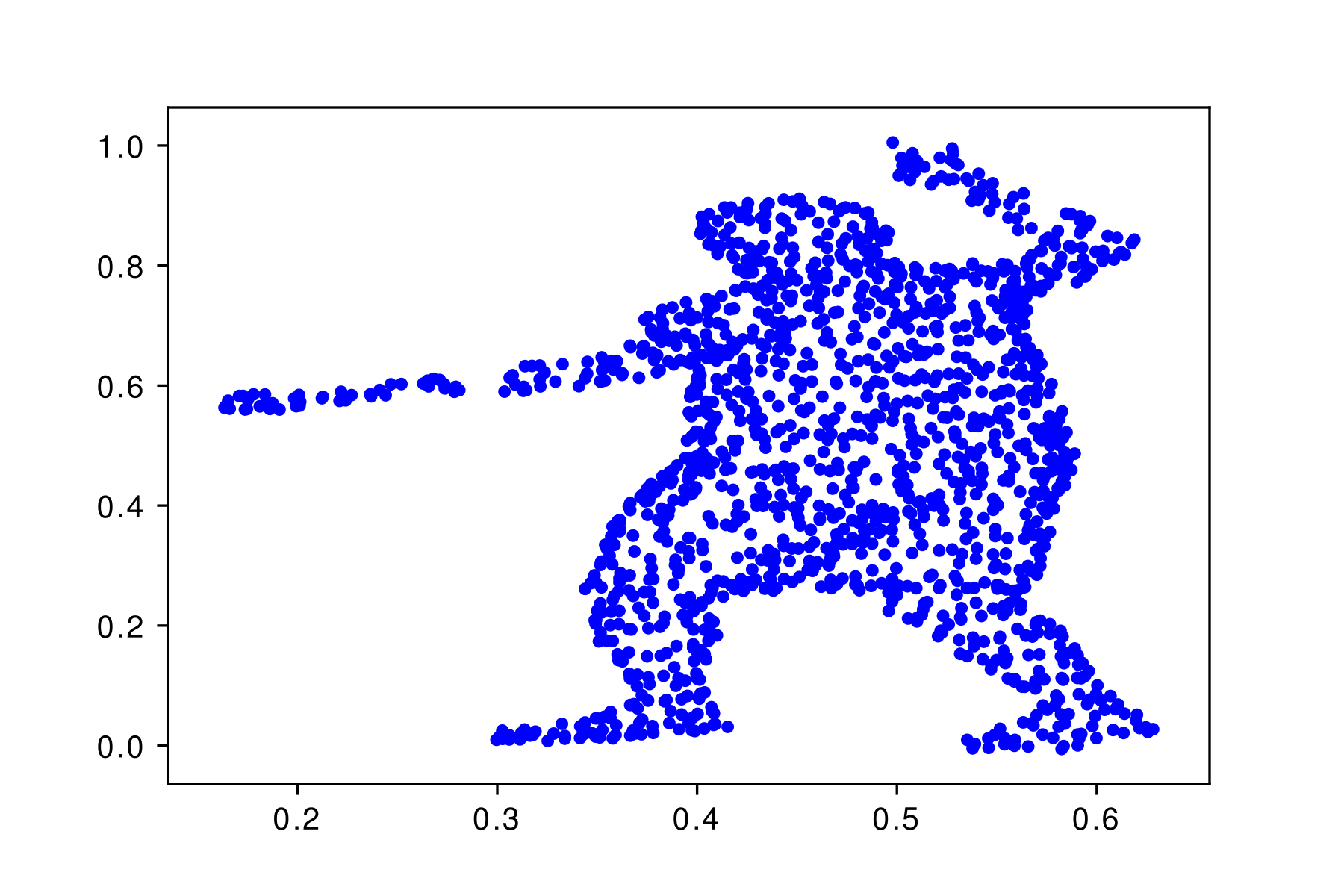}
		\caption{}
		\label{fig:prop_kipf}
	\end{subfigure}%
	~
	\begin{subfigure}[t]{0.20\textwidth}
		\includegraphics[width=1.5in, height=1.5in]{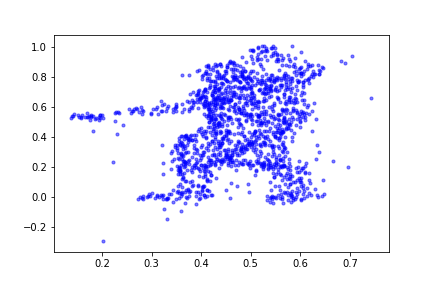}
		\caption{}
		\label{fig:LSTM_Greg}
	\end{subfigure}% 
	\begin{subfigure}[t]{0.20\textwidth}
		\includegraphics[width=1.5in, height=1.5in]{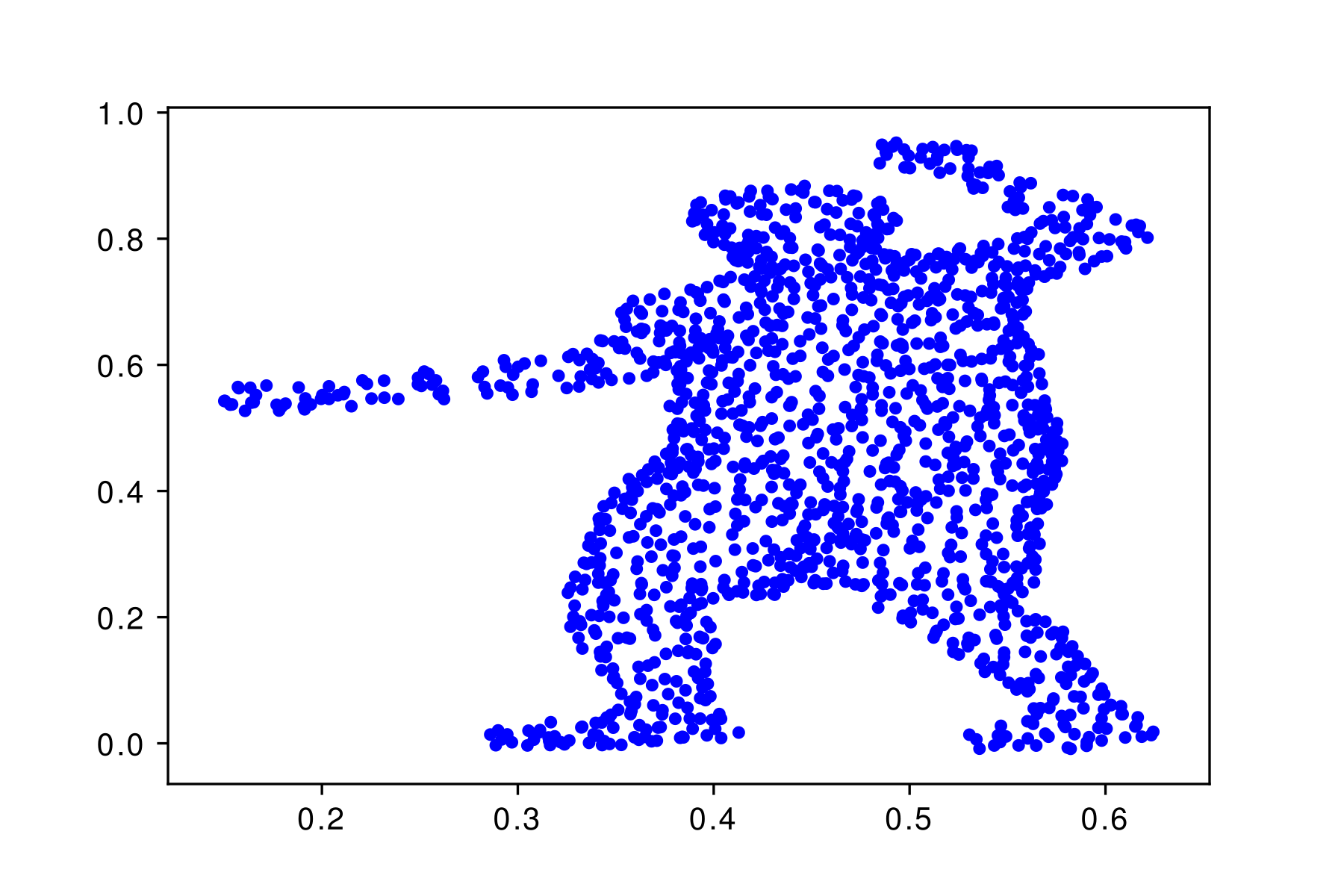}
		\caption{}
		\label{fig:LSTM_GCN}
	\end{subfigure}% 
	%    \begin{subfigure}[t]{0.35\textwidth}
	%        \includegraphics[width=2.5in,height=2.5in]{Results_NonDeepImgIndex111}
	%        \caption{Non-deep model}
	%        \label{fig:Non-deep}
	%    \end{subfigure}% 
	\caption{\footnotesize{\emph{Dynamic 3D point cloud dataset}. The colored dot indicates a node of the 3D point cloud data. Each image shows the predicted 3D point cloud frame using different architectures.} (a). Ground truth (b). Proposed model (filter based on~\cite{defferrard2016convolutional}) (c). Proposed model (filter based on ~\cite{ThomasKipf}) (d). LSTM with graph regularizer (e). LSTM with GCN~\cite{Seo_2018}.}
	
	\label{fig:images}
	\vskip-4mm
\end{figure*} 

\begin{figure*}[!h]
	\centering
	\psfrag{celcius}{\footnotesize $^\circ$ C}
	\psfrag{5}{\tiny 5} \psfrag{6}{\tiny 6} \psfrag{7}{\tiny 7} \psfrag{8}{\tiny 8} \psfrag{9}{\tiny 9} \psfrag{10}{\tiny 10} \psfrag{11}{\tiny 11}
	\begin{subfigure}[t]{0.5\textwidth}
		\includegraphics[width=\columnwidth, height=2.0in]{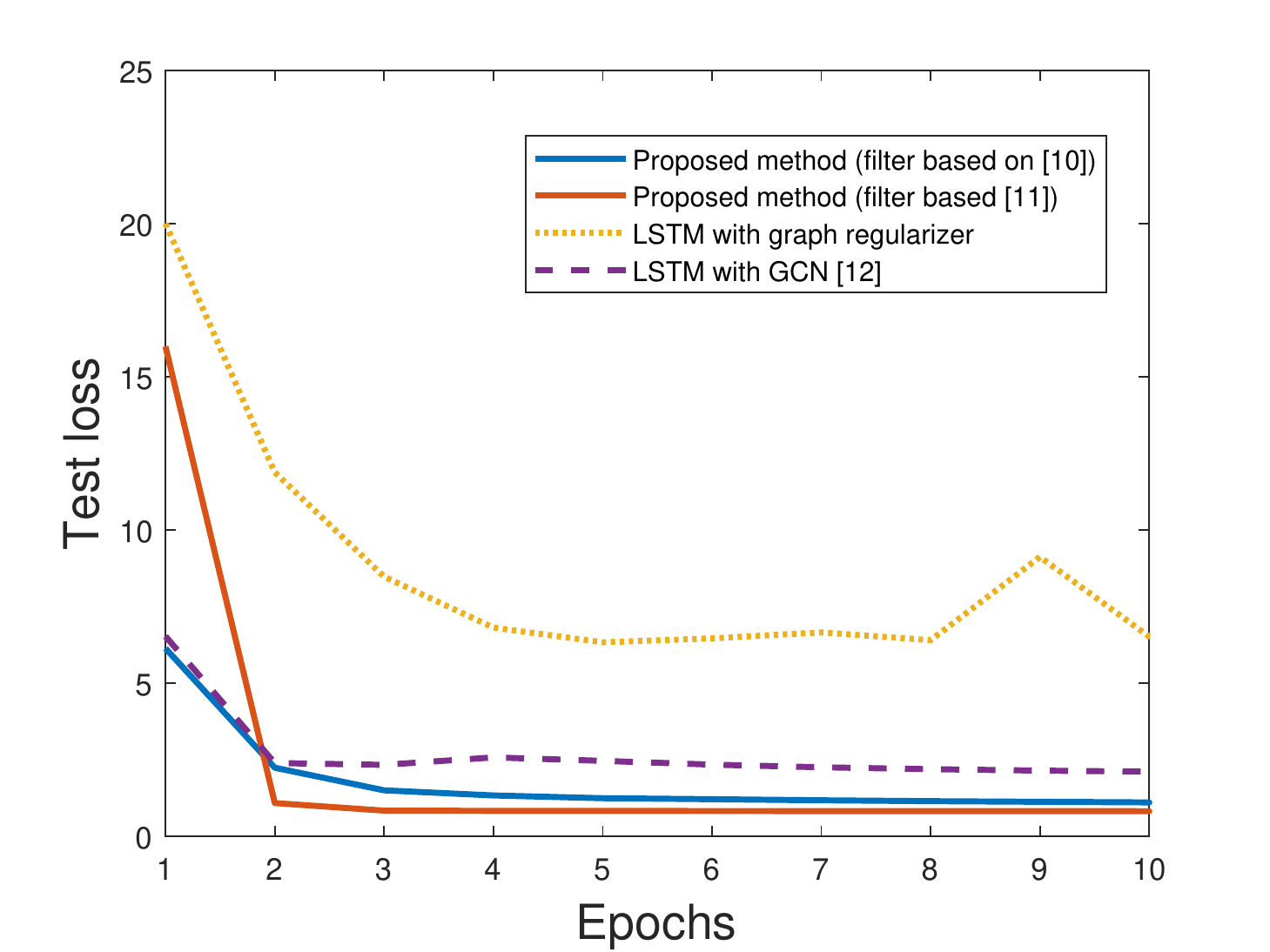}
		\caption{Test loss}
		\label{fig:Test_loss_methods}
	\end{subfigure}%
	~
	\begin{subfigure}[t]{0.5\textwidth}
		\includegraphics[width=\columnwidth,height=2.0in]{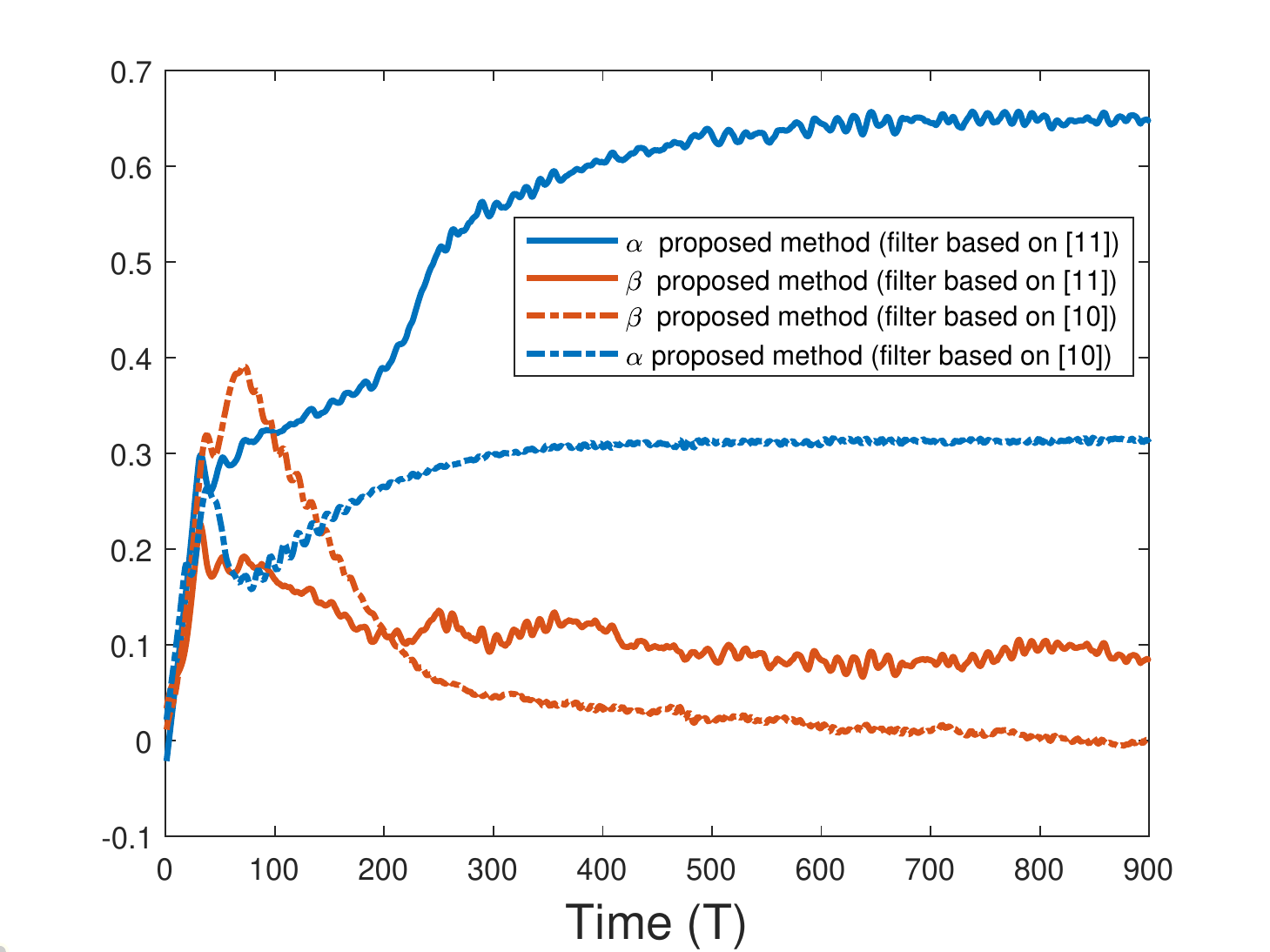}
		\caption{Variations of $\alpha$ and $\beta$}
		\label{fig:params}
	\end{subfigure}%       
	\caption{\footnotesize{\emph{Test loss of the 3D point cloud dataset}: Left: Test loss of the various architectures trained on the 3D point cloud dataset. Right: Variations of the training parameters $\alpha$ and $\beta$ with time $T$.   }}
	\label{fig:Test_loss}
	\vskip-4mm
\end{figure*} 
To analyze the problem of the vanishing and exploding gradients in the standard RNN combined with the graph convolution operation, let us assume that all the training parameters are scalars and the activation function is $\rm ReLU$. Then the term $ \prod_{t=2}^{T} \frac{\partial \bbh_t}{\partial \bbh_{t-1}}$ simplifies to
\begin{equation}\label{eq:S_RNN_cond}
\begin{aligned}
\prod_{t=2}^{T} \frac{\partial \bbh_t}{\partial \bbh_{t-1}} =  (u\bbL)^{T-2}.
\end{aligned}
\end{equation}
The stability of the gradient depends on the largest eigenvalue of \eqref{eq:S_RNN_cond}.
If its value is sufficiently small, $\rm{i.e.,} < 1$ the gradient will shrink exponentially. Moreover, if its value is large, the gradient will explode.  
For $\tilde{\bbL}$, the eigenvalues in the range $[0,2]$, thus gradient for the standard RNN, will be an exponential in $T$. This implies that, relative to the largest eigenvalue, the gradient may explode or vanish exponentially, leading to numerical instabilities during training.

In order to stabilize the gradients, we add a weighted residual connection, with two trainable parameters $\alpha$ and $\beta$ as in \eqref{FRNN} for which \eqref{eq:S_RNN_cond} now becomes $ \prod_{t=2}^{T} \frac{\partial \bbh_t}{\partial \bbh_{t-1}} = (\alpha\bbU\bbL+ \beta \bbI_N)^{T-2}$. 
By appropriately choosing $\alpha$ and $\beta$, the gradients may be stabilized. 
More generally, the condition number of $ \prod_{t=2}^{T} \frac{\partial \bbh_t}{\partial \bbh_{t-1}}$, which is bounded as
\begin{equation}\label{eq:cond_Num}
\begin{aligned}
M \leq \dfrac{(1+\frac{\alpha}{\beta}\max_t\|\bbD_t \bbU \bbL\|^2_F )^{T-2}}{(1-\frac{\alpha}{\beta}\max_t\| \bbD_t \bbU. \bbL\|^2_F )^{T-2}}
\end{aligned}
\end{equation}  
This determines the stability of the gradient.
In contrast to the standard RNN, if $\beta = 1 $ and $\alpha = 0$, then the condition number of $M $ is bounded $1$. Thus leading to stable gradients during training, but completely ignores the training data.
So, we allow the parameters $\alpha$ and $\beta $ to be trained such that FGRNN updates the hidden states in a controlled manner.
In other words, the parameters $\alpha $ and $\beta $ limit the extent to which the current feature vector $\bbx_t$ updates the hidden states $\bbh_t$. 
For instance, if $\beta \approx T \max_t\| \bbD_t \bbU \bbL\|^2_F$ and $\alpha \approx 1-\beta $, then $M = \ccalO(1)$.
Also, the FGRNN controls the condition number of the gradient using only two additional parameters as compared to the standard RNN.
Furthermore, the existing unitary RNN methods for the data defined on a regular domain are motivated by a similar observation, where they control the stability of the gradients by restricting $\bbU$ to a unitary matrix. 
However, while dealing with graph data, $\bbD_t \bbU\bbL$ might still become ill-conditioned even with an unitary $\bbU$.
Thus they might still have the vanishing gradient problem.
The proposed method allows the residual weights $\alpha$ and $\beta$ to be trained such that the condition number $M$ is restricted and thus prevents the numerical instabilities due to vanishing and exploding gradients.
%\vspace*{-0.0755in}
\section{Numerical analysis}
In this section, we test the proposed FGRNN architecture and compare its performance with the traditional architectures that combine GCNs with a stable variant of RNN, namely, LSTM.
We compare the proposed method with $(1)$. the baseline method that combines GCN with LSTM~\cite{Seo_2018}, $(2)$. LSTM with a graph regularizer in the loss function, $(3)$. proposed method, i.e., FGRNN with different convolution operators, i.e., Chebyshev polynomial \eqref{eq:Conv_1} and its first order approximation \eqref{eq:Conv_2}. 

We use the dynamic 3D point cloud dataset of a human pose. 
Each human pose is captured using $1502$ 3D data points and there are $573$ such frames.
This corresponds to a graph data, where the data points correspond to the 3D displacement of the nodes of the underlying graph.
Each data frame at time instance $t$ is given by $\bbX_t \in \reals^{N \times 3}$ where $N = 1502$ is the number of nodes.
We use $80 \%$ of the frames for training the network and the remaining $20\%$ of the time frames to test the performance.

For this dataset, we are interested in predicting the most likely next 3D point frame (here the next human pose) given the previous $T$ frames of data as in \eqref{eq:prediction_eqn}, where $P(\bbX_{t+1}|\bbX_{t-T},...,\bbX_t )$ models the probability of the frame $\bbX_{t+1}$ given the past $T$ observations. 
As the task we are interested in is the prediction, we define the loss occured for predicting the next data frame at time $t$ as $\|\bbX_{t+1} - \hat{\bbX}_{t+1}\|^2_F$,  where each $\bbX_{t+1}$ is a 3D data point. 
All the architectures (except LSTM with a graph regularizer) are trained by minimizing the aforementioned loss function using BPTT. 
For LSTM with a graph regularizer architecture, we define the loss function as $\|\bbX_{t+1} - \hat{\bbX}_{t+1}\|^2_F + \lambda {\rm{tr}} (\bbX_{t+1}^T\bbL\bbX_{t+1})$, where $\lambda$ is a positive regularizer chosen based on a grid search that leads to the best test loss.
We construct the Laplacian matrix $\bbL$ from the training data using K-nearest-neighbor such that each node has a degree of $6$.
All the architectures are trained using ADAM optimizer with a learning rate of $10^{-2}$ and a decay rate of $0.9$.
All the models are implemented in Tensorflow$:{\rmr} 1.15$ and each model is run for $10$ epochs and their test loss is shown in Fig. \ref{fig:Test_loss_methods}.

For a fair comparison with~\cite{SHI}, all the experimental models are implemented with the model defined in \eqref{eq:FGRNN_equn}, which is the same as replacing the 2D convolutions by graph convolutions.
Fig. \ref{fig:Test_loss} shows the performance of the various models implemented, and we observe that all the models converge before $10$ epochs.
These results show the ability of the proposed method to capture the structure in the graph time-series. We can observe from Fig. \ref{fig:Test_loss_methods} that FGCNNs implemented with graph filter based on~\cite{ThomasKipf} and~\cite{defferrard2016convolutional} offer better performance than regular LSTMs with a graph regularizer and architecture that combines GCN with LSTM~\cite{SHI}.
In Fig. \ref{fig:images}, we illustrate the 3D point cloud frames of the ground truth, and the estimated frames from various architectures we have considered.
We see that the predicted 3D point cloud frame by the proposed method is more consistent with the ground truth than the frame estimated by LSTM with a graph regularizer.  Moreover, the predicted feature is visually the same as the baseline methods (namely, LSTMs with a graph convolutions). 
This demonstrates that the problem of vanishing and exploding gradients can be overcome by the addition of a simple weighted residual connection to the standard RNN, which means that the proposed FGRNN is stable and can be trained efficiently.

Table \ref{tab:my-table} shows the computational complexity in terms of the number of training parameters for the different considered methods. We can see that the proposed method is computationally efficient than any other baseline methods as the number of trainable parameters in the proposed architecture is the least. This demonstrates that the proposed FGRNN models are accurate and faster to train as compared to the baseline models.
\begin{table}[]
	\begin{tabular}{|l|l|l|}
		\hline
		& \# \textbf{parameters} & \textbf{point cloud} \\ \hline
		Standard FRNN    & $3 N^2 + 2N +2$ & 6771018 \\ \hline
		Proposed (filter based on~\cite{defferrard2016convolutional})    & $3K +2N +2$ & \textbf{3015} \\ \hline
		Proposed (filter based on~\cite{ThomasKipf}) & $3P^2+2N+2$ &\textbf{3033} \\ \hline
		LSTM without GCN & $8N^2+4N$ &18054040\\ \hline
		LSTM with GCN ~\cite{Seo_2018} & $4N+ 8K $ & 6032\\ \hline
	\end{tabular}
	\caption{Comparison between the models in terms of number of trainable parameters.}
	\label{tab:my-table}
\end{table}
Finally, Fig. \ref{fig:params} shows the learnt $\alpha$ and $\beta$ on the datasets with $T$ time steps. It is clear from the figure that the learned $\beta$ is a decreasing function of $T$. Moreover, $\alpha$ can be seen close to $1-\beta$ for large $T$, while corroborates the FGRNNs theoretical analysis.

\section{Conclusions}
This paper proposes a Fast Graph Recurrent Neural Network (FGRNN) architecture for efficient training and prediction of data defined on irregular (non-euclidean) domains.
The sandard RNN architecture is known to be unstable during training due to vanishing and exploding gradients. 
Hence, gated architectures, namely, LSTMs and GRUs, are proposed to alleviate this issue at the cost of the computational complexity.
FGRNN architecture is obtained by incorporating a weighted residual connection with only two scalar parameters into the standard RNN architecture.
FGRNN model has a fewer number of training parameters, lesser training times, and is more stable than the standard RNN.
These architectures can match the state-of-the-art gated RNN architectures with a significantly lower number of training parameters and computational cost.

%\pagebreak

\bibliographystyle{IEEEtran}
{\footnotesize
\bibliography{IEEEabrv,sample}}

\begin{thebibliography}{10}
\providecommand{\url}[1]{#1}
\csname url@samestyle\endcsname
\providecommand{\newblock}{\relax}
\providecommand{\bibinfo}[2]{#2}
\providecommand{\BIBentrySTDinterwordspacing}{\spaceskip=0pt\relax}
\providecommand{\BIBentryALTinterwordstretchfactor}{4}
\providecommand{\BIBentryALTinterwordspacing}{\spaceskip=\fontdimen2\font plus
\BIBentryALTinterwordstretchfactor\fontdimen3\font minus
  \fontdimen4\font\relax}
\providecommand{\BIBforeignlanguage}[2]{{%
\expandafter\ifx\csname l@#1\endcsname\relax
\typeout{** WARNING: IEEEtran.bst: No hyphenation pattern has been}%
\typeout{** loaded for the language `#1'. Using the pattern for}%
\typeout{** the default language instead.}%
\else
\language=\csname l@#1\endcsname
\fi
#2}}
\providecommand{\BIBdecl}{\relax}
\BIBdecl

\bibitem{Donahue:2017:LRC:3069214.3069251}
J.~Donahue, L.~A. Hendricks, M.~Rohrbach, S.~Venugopalan, S.~Guadarrama,
  K.~Saenko, and T.~Darrell, ``Long-term recurrent convolutional networks for
  visual recognition and description,'' \emph{IEEE Trans. Pattern Anal. Mach.
  Intell.}, vol.~39, no.~4, pp. 677--691, Apr. 2017.

\bibitem{Karpathy}
A.~{Karpathy} and L.~{Fei-Fei}, ``Deep visual-semantic alignments for
  generating image descriptions,'' \emph{IEEE Trans. on Pattern Analysis and
  Machine Intelligence}, vol.~39, no.~4, pp. 664--676, Apr. 2017.

\bibitem{Vinyals_2015}
O.~Vinyals, A.~Toshev, S.~Bengio, and D.~Erhan, ``Show and tell: A neural image
  caption generator,'' \emph{IEEE Conference on Computer Vision and Pattern
  Recognition (CVPR)}, Jun. 2015.

\bibitem{GeometricDeepLearning}
M.~M. {Bronstein}, J.~{Bruna}, Y.~{LeCun}, A.~{Szlam}, and P.~{Vandergheynst},
  ``Geometric deep learning: Going beyond euclidean data,'' \emph{IEEE Signal
  Processing Magazine}, vol.~34, no.~4, pp. 18--42, July 2017.

\bibitem{monti2017geometric}
F.~Monti, M.~Bronstein, and X.~Bresson, ``Geometric matrix completion with
  recurrent multi-graph neural networks,'' in \emph{Advances in Neural
  Information Processing Systems}, Sep. 2017, pp. 3697--3707.

\bibitem{Bigdata}
A.~{Sandryhaila} and J.~M.~F. {Moura}, ``Big data analysis with signal
  processing on graphs: Representation and processing of massive data sets with
  irregular structure,'' \emph{IEEE Signal Processing Magazine}, vol.~31,
  no.~5, pp. 80--90, Sep. 2014.

\bibitem{sandryhaila2014discrete}
A.~Sandryhaila and J.~M. Moura, ``Discrete signal processing on graphs:
  Frequency analysis.'' \emph{{IEEE} Trans. Signal Process.}, vol.~62, no.~12,
  pp. 3042--3054, 2014.

\bibitem{LearnGraphData}
X.~{Dong}, D.~{Thanou}, M.~{Rabbat}, and P.~{Frossard}, ``Learning graphs from
  data: A signal representation perspective,'' \emph{IEEE Signal Processing
  Magazine}, vol.~36, no.~3, pp. 44--63, May 2019.

\bibitem{chepuri2016learning}
S.~P. Chepuri, S.~Liu, G.~Leus, and A.~O. Hero~III, ``Learning sparse graphs
  under smoothness prior,'' in \emph{Proc. of the IEEE International Conference
  on Acoustics, Speech, and Signal Processing (ICASSP)}, New Orleans, USA, Mar.
  2017.

\bibitem{hochreiter1997long}
S.~Hochreiter and J.~Schmidhuber, ``Long short-term memory,'' \emph{Neural
  computation}, vol.~9, no.~8, pp. 1735--1780, 1997.

\bibitem{defferrard2016convolutional}
M.~Defferrard, X.~Bresson, and P.~Vandergheynst, ``Convolutional neural
  networks on graphs with fast localized spectral filtering,'' in
  \emph{Advances in neural information processing systems}, 2016.

\bibitem{ThomasKipf}
T.~N. Kipf and M.~Welling, ``Semi-supervised classification with graph
  convolutional networks,'' in \emph{International Conference on Learning
  Representations (ICLR)}, 2017.

\bibitem{Seo_2018}
Y.~Seo, M.~Defferrard, P.~Vandergheynst, and X.~Bresson, ``Structured sequence
  modeling with graph convolutional recurrent networks,'' \emph{Lecture Notes
  in Computer Science}, p. 362–373, 2018.

\bibitem{kusupati2018fastgrnn}
A.~Kusupati, M.~Singh, K.~Bhatia, A.~Kumar, P.~Jain, and M.~Varma, ``Fastgrnn:
  A fast, accurate, stable and tiny kilobyte sized gated recurrent neural
  network,'' in \emph{Advances in Neural Information Processing Systems}, 2018.

\bibitem{Bruna}
J.~Bruna, W.~Zaremba, A.~Szlam, and Y.~Lecun, ``Spectral networks and locally
  connected networks on graphs,'' Dec. 2013.

\bibitem{SHI}
S.~Xingjian, Z.~Chen, H.~Wang, D.-Y. Yeung, W.-K. Wong, and W.-c. Woo,
  ``Convolutional lstm network: A machine learning approach for precipitation
  nowcasting,'' in \emph{Advances in neural information processing systems},
  2015.

\end{thebibliography}
%\fontsize{0.5pt}{1pt}

\end{document}